\documentclass[english,prb,aps,twocolumn,floatfix,superscriptaddress]{revtex4-1}

\usepackage[linktocpage,bookmarksopen,bookmarksnumbered]{hyperref}
\usepackage{graphicx}
\usepackage{dcolumn}
\usepackage{amsmath,graphics,epsfig,color,verbatim,ulem}
\usepackage{amssymb}
\usepackage{babel}

\begin{document}

\title{Density functional versus spin-density  functional and the choice of correlated subspace in multi-variable effective action theories of electronic structure}

\author{Hyowon Park}
\thanks{Present Address: Department of Physics, University of Illinois at Chicago, Chicago, IL 60607, USA}
\email{hyowon@uic.edu}
\affiliation{Department of Applied Physics and Applied Mathematics, Columbia University, New York, NY 10027, USA}
\affiliation{Department of Physics, Columbia University, New York, NY 10027, USA}
\author{Andrew J. Millis}
\affiliation{Department of Physics, Columbia University, New York, NY 10027, USA}
\author{Chris A. Marianetti}
\affiliation{Department of Applied Physics and Applied Mathematics, Columbia University, New York, NY 10027, USA}

\date{\today}

\begin{abstract}

Modern extensions of density functional theory  such as the density functional theory plus  U and the density functional theory plus dynamical mean-field theory require choices, including selection of variable (charge vs spin density) for the density functional and specification of the correlated subspace.  This paper examines these issues in the context of the ``plus U'' extensions of density functional theory, in which additional correlations on specified correlated orbitals are treated using a Hartree-Fock approximation.  Differences between using  charge-only or spin-density-dependent exchange-correlation functionals and between Wannier and projector-based definitions of the correlated orbitals are considered on the formal level and in the context of the structural energetics of the  rare earth nickelates.  It is demonstrated that theories based on spin-dependent  exchange-correlation functionals can lead to large and in some cases unphysical effective on-site exchange couplings.  Wannier and projector-based definitions of the correlated orbitals lead to  similar behavior near ambient pressure, but substantial differences are observed at large pressures.  Implications for other beyond density functional methods such as the combination of density functional and dynamical mean field theory are discussed. 

\end{abstract}

\maketitle

\section{Introduction}
Modern theories of electronic structure can be  formally constructed in terms of functionals of observables of interest whose stationary points deliver the values of the observables\cite{Kotliar:06}.  Practical use of this formal construction requires a choice of variables and of approximations to the functional.  Perhaps the most common choice is density functional theory\cite{Hohenberg1964864,Kohn19651133} (DFT), which can be formulated as an effective action that is a  functional only of the electron density\cite{Fukuda1994833} (ie. not spin resolved). While DFT is in principle exact, existing approximations have had difficulty capturing phenomena related to the formation of local magnetic moments. For example, neither the local density approximation\cite{Kohn19651133} (LDA) nor the generalized gradient approximation (GGA)\cite{Ma196818} provide correct accounts of the structural energetics of  layered and spinel manganites that exhibit cooperative Jahn-Teller distortions associated with the high spin state of Mn$^{3+}$,~\cite{Van_der_ven200021} because at ambient pressure both LDA and GGA incorrectly predict that the Mn ion is in a nominal $|t_{2g}^4 e_g^0 \rangle$ low-spin configuration instead of  the proper high-spin $|t_{2g}^3 e_g^1 \rangle$ state.  Indeed it seems intuitively clear that it would be prohibitively difficult to construct a functional based only on the density that could capture this sort of effect, while a functional of the spin-density might have a robust approximation which can capture this physics. Additionally, a functional of the spin-density will clearly allow predictions to be made about magnetism, and this avenue has been pursued since the inception of DFT\cite{Kohn19651133,Vonbarth19721629}.

Functionals of both the charge and spin density such as the the local spin-density approximation (LSDA)\cite{Vonbarth19721629,Rajagopal19731912,Gunnarsson19764274} and the spin-dependent generalized gradient approximation (SGGA)\cite{Rasolt19773234,Rasolt198145} have been constructed.  Such theories are often refered to as density functional theories, but in this paper we strictly distinguish terms, using the term density functional theory (DFT) to refer to theories such as the LDA and GGA that are based on  a functional of the density only, and referring to theories such as the local spin density approximation (LSDA) or the spin-dependent GGA (SGGA) as spin-density functional theories (SDFT).

SDFT theories perform far better than DFT theories in describing the energetics of magnetic insulators, resolving, for example, the problems with manganites noted above.\cite{Van_der_ven200021} SDFT theories additionally  make predictions about  spin magnitudes and the nature of ordered states.  However,  the known implementations of SDFT fail to correctly describe many aspects of the physics and structure of strongly correlated electron systems,  for example providing  qualitatively incorrect  structures for the rare-earth nickelates\cite{Park:14,Park2014235103} (see, e.g.  Ref. \onlinecite{Kotliar:06} for additional examples).

These difficulties motivated  the construction of new effective action theories that depend not only on the density or the spin-density,  but also on additional properties of a subspace of orbitals for which  correlations are believed to be relevant\cite{Anisimov1991943,Kotliar:06}. Subspaces which have been treated in this way include the transition metal $d$ orbitals in transition metal oxides and the lanthanide/actinide $f$ levels in heavy fermion compounds.   Various different variables can be defined from the subspace of correlated orbitals. In this paper  we focus on the historically important and currently widely used choice of the site-local spin and orbitally resolved density matrix associated with the correlated subspace\cite{Anisimov1997767}; however, we expect that our findings are relevant to other variable choices; in particular to the case of dynamical mean field theory where the additional variables are the components of the  local Green's function.  A straightforward functional to use with the local spin-resolved density matrix is the Hartree-Fock energy functional, defined using site-local matrix elements of the Coulomb interaction. The functional resulting from this combination of a local  Hartree-Fock functional of a correlated subspace  and a standard functional of the density or spin-density is commonly  referred to as a ``+U'' extension of density functional theory. 

The paper that introduced the ``+U" approach\cite{Anisimov1991943}  employed a functional of the electron density only, so is referred to here as a DFT+U approach. However, the vast majority of  subsequent papers\cite{Anisimov1997767} employ SDFT functionals and are referred to here as SDFT+U approaches. Despite the many successes of the +U methodology, basic points including the rationale for choosing SDFT+U in preference to DFT+U  and the factors influencing the construction of the correlated subspace have not been clearly discussed.  While the general formalism may be applied with any choice of orbital, it is important in practice to choose orbitals that are optimal in the context of the other approximations used in constructing the theory.  While there is no clear prescription for  doing this, the local nature of the approximations which will ultimately be used suggests that it is sensible to choose the correlated subspace to consist of well localized `atomic-like' orbitals.  These orbitals may be constructed from by projecting onto a set of localized orbitals defined to lie within atomic spheres. This choice is natural, given that many basis sets for electronic structure already utilize projectors. Projectors are used in various beyond-DFT methods including (S)DFT+U~\cite{Bengone:00} as well as DFT+DMFT~\cite{Anisimov:97,Lichtenstein:98,Savrasov:04,Pourovskii:07,Wills:12,Haule:10,Aichhorn:09}.  Alternatively, Wannier functions may  be used to construct the correlated orbital sets for beyond DFT calculations. Various forms of the Wannier function have been used for DFT+U~\cite{Anderson:00} and DFT+DMFT~\cite{Pavarini:04,Anisimov:05,Korotin:08,Amadon:08,Amadon:12,Dai:12,Lechermann:06,Park:12} including the projected Wannier function, $N$th-order muffin-tin orbitals~\cite{Anderson:00}, and the maximally localized Wannier function (MLWF)~\cite{Marzari:97,Marzari:12,Wannier}.

In this paper we  describe the physical differences between DFT+U and SDFT+U and  provide guidelines to enable researchers to choose between them. We further compare the effect of different correlated orbitals sets (Projector vs Wannier)  on  energy calculations within the (S)DFT+U method. We also present a comprehensive discussion of the issues arising when the +U methodologies are combined with the Projector Augmented Plane Wave (PAW) formalism widely used to perform efficient (S)DFT calculations.  In addition to the formalism we  provide a quantitative application in the context of  the relation between crystal structure and  energetics of the rare-earth nickelates. This family of materials provides a useful  benchmark because its members exhibit a structural phase transition which is not correctly captured either by DFT or by  SDFT calculations. We compare DFT+U and SDFT+U, using both projectors and Wannier functions to construct the correlated subspace. These results can also be directly compared to our recent DFT+DMFT total energy calculations for the same class of materials\cite{Park2014235103}.  

We note in passing that an additional important issue in DFT+U and SDFT+U theories is the so-called double counting correction, introduced to account for the fact that the local interactions denoted by $U$ and $J$ are to some degree present already in the (S)DFT. The double counting issue has been addressed in great detail in previous work \cite{Park2014235103} and is not critical to the issues examined here. Therefore in this paper we use the conventional definition of the ``fully localized-limit" double-counting \cite{Anisimov:93,Sawatzky:94}.

The rest of this paper is organized as follows. Section ~\ref{sec:Theory} presents the basic formalism. Section ~\ref{sec:PAW} provides a careful discussion of the issues involved in combining the +U formalism with the projector augmented plane wave method. Section ~\ref{sec:Theory-3} presents expressions for forces, needed in optimizing structures.  Section ~\ref{sec:Theory-4} compares the DFT+U and SDFT+U (with projector and Wannier definitions of the correlated subspace) predictions for the structural properties of the rare earth nickelates as a function of unit cell volume, while section ~\ref{sec:Results} provides a comparison of predictions for the rare earth nickelate phase diagram and the equilibrium volume. Section ~\ref{sec:conclusion} contains a summary and conclusions.  Relevant computational details are given in Appendix ~\ref{Appendix:sec:Theory-4}.

\section{Formalism}
\label{sec:Theory}

In this section we present explicit formulae for the energy functional, using a variant of presentation of the DFT+DMFT functional given in Ref.~\onlinecite{Park2014235103}.  We derive the total energy functional for SDFT+U, and then obtain the DFT+U functional as a special case. 

The SDFT+U total energy functional is defined by  
%
\begin{align} \nonumber
E[\rho^{\sigma},n^{\tau\sigma}] &= \mathbf{Tr}[\langle\hat{H}^{\sigma}_{U}\rangle]
-\mathbf{Tr}[\hat{V}^{\sigma}_{Hxc}\cdot\rho^{\sigma}]
\\
& - \mathbf{Tr}[\hat{V}^{\sigma}_{int}\cdot n^{\tau\sigma}]  +
E^{Hxc}[\rho^{\sigma}] +
 E^{int}[n^{\tau\sigma}]
\label{eq:sdftu}
\end{align}
where $\rho^{\sigma}$ denotes the charge density of electrons  with  spin $\sigma$ (we neglect spin-orbit coupling here for simplicity) and $n^{\tau\sigma}$ is a density matrix within the correlated subspace of an atom $\tau$.
Here the bracket $\langle\rangle$ means that the eigenstates of $\hat{H}^{\sigma}_{U}$ are summed over for the eigenvalues less than the Fermi energy.  


The functional $E^{Hxc}$ is the familiar Hartree and exchange-correlation energy functional of the DFT theory to be used and  $\hat{V}^{\sigma}_{Hxc}=\delta E^{Hxc}/\delta \rho^{\sigma}$ is the corresponding Hartree-exchange-correlation potential.

$E^{int}$ is the  combination of a Hartree-Fock potential energy $E^{pot}$ defined within  the correlated subspace  and a double counting correction $E^{DC}$ introduced to remove from this potential the parts already included in the underlying DFT:
\begin{equation}
E^{int}=E^{pot}-E^{DC}.
\label{Eintdef}
\end{equation}

In the applications presented here we follow the common practice in the literature by choosing the correlated subspace to be particular orbitals $m$ of spin $\sigma$ on particular atoms $\tau$ in the solid; for example the $3d$ orbitals in a first row transition metal ion or the $4f$ orbitals in a lanthanide ion, so that 
\begin{equation}
E^{pot} =\!\!\!\!\!\!\!\! \sum_{\tau,m_1m_2\atop m_3m_4  \sigma_1\sigma_2}\!\!\!\!\!\!\!\! (U_{m_1m_3m_2m_4}^{\sigma_1\sigma_2\sigma_1\sigma_2}-U_{m_1m_3m_4m_2}^{\sigma_1\sigma_2\sigma_2\sigma_1}\cdot \delta_{\sigma_1\sigma_2})n^{\tau\sigma_1}_{m_1m_2}n^{\tau\sigma_2}_{m_3m_4}
\label{eq:E_pot}
\end{equation}
where  the $U_{m_1m_3m_2m_4}^{\sigma_1\sigma_2\sigma_1\sigma_2}$ are the site-local matrix elements of the Coulomb interaction within the correlated subspace, appropriately renormalized by the solid-state environment.  This paper will present an application of the formalism to the case of transition metal $d$-orbitals where  (if spin orbit coupling is neglected), the $U_{m_1m_3m_2m_4}^{\sigma_1\sigma_2\sigma_1\sigma_2}$ may be parametrized by the Slater integrals $F^0$, $F^2$, and $F^4$. We typically further assume that the $d$-wave functions are sufficiently similar to their free-space forms that $F^4=0.625F^2$. It is then   conventional to define  on-site interaction $U$ and the Hund's coupling $J$ via $F^0=U$,  $F^2=(14/1.625)J$.~\cite{Anisimov:93}

The double counting energy $E^{DC}$ is the subject of continuing discussion in the literature but these subtleties are not relevant here. We  adopt the `fully localized limit' form given by
\begin{eqnarray}
E^{DC}[N_\tau^{\sigma}]&=&\frac{U}{2}N_\tau(N_\tau-1)-\frac{J}{2}\sum_{\sigma}N_\tau^{\sigma}(N_\tau^{\sigma}-1),
\label{eq:E_DC} 
\end{eqnarray}
where $N_\tau^\sigma$ is the total number of electrons of spin $\sigma$ on  site $\tau$ and $N_\tau=\sum_\sigma N_\tau^{\sigma}$ is the total on-site charge density in the correlated subspace.  Note that use of SDFT (in other words  the dependence of  $E^{Hxc}[\rho^{\sigma}]$ on the spin density) means that   that the double counting potential $E^{\sigma}_{DC}$ in Eq.$\:$\ref{eq:E_DC} also depends on spin indices. 

The corresponding interaction potential  $\hat{V}^{\sigma}_{int}$ is obtained as $V^{\sigma}_{int}=\partial E^{int}/\partial n^{\tau \sigma}$:
\begin{equation}
\hat{V}^{\sigma}_{int}= \sum_{m,m'}|\tau,m,\sigma\rangle (V_{pot,m m'}^{\tau\sigma}-V^{\tau\sigma}_{DC}) \langle\tau,m',\sigma|,
\label{eq:E_int}
\end{equation}
where
\begin{equation}
V_{pot,m m'}^{\tau\sigma} = \sum_{m_1m_2\sigma_1} (U_{m_1mm_2m'}-U_{m_1mm'm_2}\cdot \delta_{\sigma_1\sigma})n^{\tau\sigma_1}_{m_1m_2}
\label{eq:V_POT} 
\end{equation}
and
\begin{eqnarray}
V^{\tau\sigma}_{DC} & = & U(N_d-\frac{1}{2})-J(N_d^{\sigma}-\frac{1}{2}).
\label{eq:V_DC}
\end{eqnarray}

Finally, the effective Hamiltonian $\hat H_U^{\sigma}$ is 
\begin{equation}\label{eq:KSD}
\hat{H}_{U}^\sigma=\hat H^\sigma_{KS} + \hat{V}_{int}^\sigma, 
\end{equation}
where the Kohn-Sham Hamiltonian $\hat{H}^{\sigma}_{KS}$ is 
\begin{equation}
\hat{H}^{\sigma}_{KS}[\rho^{\sigma},\mathbf{R}]=-\frac{1}{2}\hat{\nabla}^2+\hat{V}_{ext}[\mathbf{R}]+\hat{V}^{\sigma}_{Hxc}[\rho^{\sigma}].
\label{eq:H_DFT}
\end{equation}
$\hat H_U^{\sigma}$ is the analogue of the Kohn-Sham Hamiltonian $\hat H^\sigma_{KS}$ of SDFT.
$\hat{V}_{ext}$ is an external potential arising from atomic nuclei at the position $\mathbf{R}$ (the interaction between nuclei is not explicitly denoted  but is included in the formalism).

The physical state of the system at zero temperature is obtained by extremizing
Eq.$\:$\ref{eq:sdftu}  with respect to variations in $\rho^{\sigma}$ and
$n^{\tau\sigma}$. To perform the extremization, we solve the eigenvalue problem of
$\hat H_U^\sigma$ and find the eigenstates $|\Psi^\sigma_{i\mathbf{k}}\rangle$
($i$ is a band index and $k$ is a momentum in the first Brillouin zone).  We
then determine the local charge density
$\rho^\sigma(\mathbf{r})=\sum_{ik}f^{\sigma}_{ik}\left|\Psi^\sigma_{ik}(\mathbf{r})\right|^2$
($f$ is the Fermi function and it is evaluated at zero temperature ($T$=0)). The on-site density matrix is also determined from
$|\Psi\rangle$ using for example the method described in Section
~\ref{sec:Theory-2-1} or Section ~\ref{sec:Theory-2-2}. Finally, we require consistency
between the $\rho^\sigma$ and $n^{\tau\sigma}$ and the Kohn-Sham and $V_{int}$
potentials they imply.   

Once self-consistent solutions of  $\rho^\sigma$ and $n^{\tau\sigma}$ are obtained, the total ground-state energy $E$ can be obtained from the value of Eq.$\:$\ref{eq:sdftu} at the stationary point of both $\rho^\sigma$ and $n^{\tau\sigma}$.
It is also useful to cast the total energy functional in Eq.$\:$\ref{eq:sdftu} is a slightly different form, both for analysis and for technical reasons specially when using Wannier functions. The SDFT+U total energy functional can be decomposed into the SDFT energy ($E^{SDFT}$), the KS energy correction ($E^{\Delta KS}$), and the interaction energy correction ($E^{int}$) (defined only within the correlated subspace) as follows:
\begin{equation}
E =  E^{SDFT}+E^{\Delta KS}+E^{int},
\label{eq:Energy}
\end{equation}
where
\begin{equation}
E^{SDFT}=\mathbf{Tr}[\langle\hat{H}^{\sigma}_{KS}\rangle]-\mathbf{Tr}[\hat{V}_{Hxc}^{\sigma}\cdot\rho^\sigma]+E^{Hxc}[\rho^\sigma]
\label{eq:E_SDFT}
\end{equation}
and
\begin{equation} 
E^{\Delta KS}  = \mathbf{Tr}[\langle\hat{H}^{\sigma}_{U}\rangle]-\mathbf{Tr}[\hat{V}^{\sigma}_{int}\cdot n^{\tau\sigma}]-\mathbf{Tr}[\langle\hat{H}^{\sigma}_{KS}\rangle].
\end{equation}

DFT+U is a special case of the  SDFT+U  in which  the exchange-correlation energy depends only  on the total density $\rho$, i.e., $\Phi_{KS}(\rho^{\sigma})\rightarrow \Phi_{KS}(\rho)$.  However, spin dependence is retained in the correlated subspace, so $\mathbf{n}^{\tau\sigma}$ is still considered to  be spin dependent. Thus the total energy functional in Eq.$\:$\ref{eq:sdftu} becomes
\begin{align} \nonumber
E[\rho,n^{\tau\sigma}] &= \mathbf{Tr}[\langle\hat{H}^{\sigma}_{U}\rangle]
-\mathbf{Tr}[\hat{V}_{Hxc}\cdot\rho] \\
& - \mathbf{Tr}[\hat{V}^{\sigma}_{int}\cdot n^{\tau\sigma}]  +
E^{Hxc}[\rho] + E^{int}[n^{\tau\sigma}],
\label{eq:dftu}
\end{align}
where
\begin{equation}
\hat{H}_{U}^\sigma=\hat H_{KS} + \hat{V}_{int}^{\sigma}. 
\end{equation}

The rest of the formalism carries through as before, except that the exchange-correlation potential now depends only on $\rho$ and therefore  the   double counting correction  is taken to be spin-independent:
\begin{equation}
E^{DC}[N_\tau]=\frac{U}{2}N_\tau(N_\tau-1)-\frac{J}{4}N_\tau(N_\tau-2),
\label{eq:dc2}
\end{equation}
implying 
\begin{equation}\label{eq:Vdc2}
V_{DC}[N_\tau] = U(N_\tau-\frac{1}{2})-\frac{J}{2}(N_\tau-1).
\end{equation}

Thus in DFT+U theories  spin dependence arises only from the properties of the correlated subspace (which affect the rest of the system via hybridization).

\section{Correlated orbitals and the implemention of (S)DFT+U using the Projector-Augmented Plane Wave Method}
\label{sec:PAW}
\subsection{Overview}

In this paper, the DFT portion of the functional (Eq.$\:$\ref{eq:E_SDFT}) is implemented using the projector augmented wave (PAW) method~\cite{Blochl:1994}. The main idea of PAW is to circumvent treating the computationally inconvenient core states by use of a  linear transformation which relates an all-electron wavefunction $|\psi\rangle$ to a pseudo wavefunction $|\tilde{\psi}\rangle$.  The transformation requires the addition of   augmentation terms which can be expanded using a projector function $|\tilde{p}\rangle$ and the resulting KS Hamiltonian contains  additional terms arising from the augmentations, but because the resulting pseudo-wavefunction is smoothly varying  computations are much more efficient.  

In the PAW method, the SDFT energy functional can be split into three terms:
\begin{equation}
E^{SDFT}=\tilde{E}[\tilde{n},\hat{n},\tilde{n}_{Zc}]+E^1[n^1,n_{Zc}]-\tilde{E}^1[\tilde{n}^1,\hat{n},\tilde{n}_{Zc}]
\label{eq:E_PAW}
\end{equation}
where $\tilde{E}$ is the pseudo-energy term, $E^1$ is the on-site all-electron energy term, and $\tilde{E}^1$ is the on-site pseudo-energy term.
$\tilde{n}$ is the pseudo charge density, $n^1$ is the on-site all-electron charge density, $\tilde{n}^1$ is the on-site pseudo charge density, $\hat{n}$ is the compensation charge between $n^1$ and $\tilde{n}^1$ such that $\tilde{n}^1+\hat{n}$ has the exact same moment as $n^1$, $\tilde{n}_{Zc}$ is the pseudized core density, and $n_{Zc}$ is the all-electron core density.

The effective $\hat{H}^{\sigma}_{KS}$ for generating the pseudo wavefunction $|\tilde{\psi}\rangle$ is now given by extremizing $E^{SDFT}$ in Eq.$\:$\ref{eq:E_PAW} with respect to the pseudized charge $\tilde{\rho}^{\sigma}$:
\begin{equation}
\hat{H}^{\sigma}_{KS}=-\frac{1}{2}\hat{\nabla}^2+\tilde{v}_{eff}+\sum_{i,j}|\tilde{p}_i\rangle(\tilde{D}_{ij}+D^1_{ij}-\tilde{D}^1_{ij})\langle\tilde{p}_j|
\label{eq:H_PAW}
\end{equation}
where $\tilde{v}_{eff}$ is the effective pseudo one-particle potential obtained using $\tilde{v}_{eff}=\frac{\partial E^{SDFT}}{\partial \tilde{\rho}^{\sigma}}$.
$\tilde{D}_{ij}$, $D^1_{ij}$, and $\tilde{D}^1_{ij}$ are potentials conjugate to the density matrix of the augmentation part $\rho_{ij}$, 
ie, $\tilde{D}_{ij}=\frac{\partial \tilde{E}}{\partial \rho_{ij}}$, $D^1_{ij}=\frac{\partial E^1}{\partial \rho_{ij}}$, and $\tilde{D}^1_{ij}=\frac{\partial \tilde{E}^1}{\partial \rho_{ij}}$.

Eq.$\:$\ref{eq:E_PAW} can be cast into a similar form as Eq.$\:$\ref{eq:E_SDFT}:
\begin{equation}
E^{SDFT}[\rho^{\sigma}]=\sum_{i\mathbf{k}}f_{i\mathbf{k}}\langle \tilde{\psi}_{i\mathbf{k}}|\hat{H}^{\sigma}_{KS}|\tilde{\psi}_{i\mathbf{k}}\rangle+E^{PAW}_{dc}[\tilde{n},\hat{n},n^1,\tilde{n}^1].
\label{eq:E_PAW2}
\end{equation}
The PAW double counting correction $E^{PAW}_{dc}$ also contains a pseudo part and an augmentation part:
\begin{equation}
E^{PAW}_{dc}[\tilde{n},\hat{n},n^1,\tilde{n}^1]=\tilde{E}_{dc}[\tilde{n},\hat{n}]+E^1_{dc}[n^1]-\tilde{E}^1_{dc}[\tilde{n}^1,\hat{n}].
\end{equation}
The PAW double counting correction $E^{PAW}_{dc}$ should not be confused with the double-counting correction $E^{DC}$ required in the interaction functional.
The derivation of the above equations and the explanation of each term are given in Ref.~\onlinecite{Kresse19991758}.

The treatment of the +U interactions in the PAW formalism depends on the prescription used to construct the correlated subspace. Accordingly, the remainder of this section is divided into two parts, one dealing with the projector formalism (subsection \ref{sec:Theory-2-1})  and one with the Wannier formalism (subsection \ref{sec:Theory-2-2}).

\subsection{Projectors: ortho-normalization}
\label{sec:Theory-2-1}

We begin with the projector method, in which the components  $n^{\tau\sigma}_{mn}$  of the correlated orbital density matrix $n$ appearing in Eq.$\:$\ref{eq:sdftu} are obtained by projecting the Kohn-Sham wavefunction $\psi$ onto  the spherical harmonics $Y_{lm}$ inside an atomic sphere centered on atom $\tau$ with the radius $r^{\tau}_c$, i.e.,
\begin{equation}
n^{\tau\sigma}_{mn} = \sum_{i\mathbf{k}}f_{i\mathbf{k}}
\langle\psi^{\sigma}_{i\mathbf{k}}|\hat{P}^{\tau}_{mn}|\psi^{\sigma}_{i\mathbf{k}}\rangle.
\label{eq:densm}
\end{equation}
Here $i$ is a band index and $\mathbf{k}$ is a wavevector in the first Brillouin zone.  $f_{i\mathbf{k}}$ is the Fermi function evaluated at $T$=0 throughout our paper and $\hat{P}^{\tau}_{mn}$ is the projector function on atom $\tau$ defined by  
\begin{equation}
\langle\mathbf{r'}|\hat{P}^{\tau}_{mn}|\mathbf{r}\rangle = Y_{ln}^*(\hat{\mathbf{r}}'_{\tau})Y_{lm}(\hat{\mathbf{r}}_{\tau})\delta(r_{\tau}-r'_{\tau})\Theta\left(r_{\tau}<r^{\tau}_c\right)
\label{eq:P}
\end{equation}
where $\mathbf{r}_{\tau}=\mathbf{r}-R_{\tau}$ is the position vector defined with respect to the atomic center $R_{\tau}$ and $\Theta(x)$ is the step function such that $\Theta(x)=1$ if $x<0$ and  $\Theta(x)=0$ if $x>0$.\cite{Bengone200016392}  Note that if the fermi function is removed from   Eq.~\ref{eq:densm} and the sum is taken over  all bands $i$ and momenta $\mathbf{k}$ then standard completeness relations imply that
\begin{equation}
O^{\tau\sigma}_{mn}\equiv \sum_{i\mathbf{k}}
\langle\psi^{\sigma}_{i\mathbf{k}}|\hat{P}^{\tau}_{mn}|\psi^{\sigma}_{i\mathbf{k}}\rangle\sim O_n\delta_{mn}
\label{eq:densm1}
\end{equation}
is a diagonal matrix, whose normalization depends on the choice $r^\tau_c$ of sphere cutoff.

Within the PAW formalism, $n^{\tau\sigma}_{mn}$ is computed from the  pseudo-wavefunctions $|\tilde{\psi}\rangle$ and a pseudo-projector $\tilde{P}^{\tau\sigma}$  as 
\begin{equation}
n^{\tau\sigma}_{mn} = \sum_{i\mathbf{k}}f_{i\mathbf{k}}
\langle\tilde{\psi}^{\sigma}_{i\mathbf{k}}|\tilde{P}^{\tau}_{mn}|\tilde{\psi}^{\sigma}_{i\mathbf{k}}\rangle.
\label{eq:PAW}
\end{equation}
The pseudo-projector is  defined in terms of an appropriate set $\left|\phi_a\right>$ of solutions to the Schroedinger equation for a reference atom in free space as
\begin{equation}
\tilde{P}^{\tau}_{mn} = \sum_{ab} |\tilde{p}_{a}\rangle\langle\phi_{a}|\hat{P}^{\tau}_{mn}|\phi_{b}\rangle\langle\tilde{p}_{b}|
\label{eq:P}
\end{equation}
where the   $|\tilde{p}\rangle$ are the PAW projector functions conjugate to the $\phi_a$.  
In practice, we use the implementation in the VASP code.  

The wave functions defined by the PAW projector process do not constitute an orthonormal set because the sum is only over a subset of states and a choice of sphere radius is made. Thus the overlap matrix
\begin{eqnarray}
O^{\tau\sigma}_{mn} & = & \sum_{i\mathbf{k}}\langle\tilde{\psi}^{\sigma}_{i\mathbf{k}}|\tilde{P}^{\tau}_{mn}|\tilde{\psi}^{\sigma}_{i\mathbf{k}}\rangle 
\label{eq:O}
\end{eqnarray}
is neither diagonal nor possessing correctly normalized eigenvalue.  To obtain a properly  ortho-normalized density matrix $\bar{n}$ within the correlated subspace  we define
\begin{eqnarray}
\overline{n}^{\tau\sigma}_{mn} & = & \sum_{m'n'}(O^{\tau\sigma})^{-1/2}_{mm'}\cdot n^{\tau\sigma}_{m'n'}\cdot (O^{\tau\sigma})^{-1/2}_{n'n}.
\label{eq:norm}
\end{eqnarray}
This procedure of the ortho-normalization of the projector function is used in some  DFT+DMFT implementations~\cite{Haule:10}.

Within the PAW formalism, the effective Hamiltonian $\hat{H}^\sigma_{U}$ can be obtained by varying the energy functional Eq.$\:$\ref{eq:sdftu} with respect to the pseudized charge $\tilde{\rho}^{\sigma}=\sum_{i\mathbf{k}}f_{i\mathbf{k}}|\tilde{\psi}^{\sigma}_{i\mathbf{k}}\rangle\langle\tilde{\psi}^{\sigma}_{i\mathbf{k}}|$:
\begin{equation}\label{energyPAW}
\hat{H}^\sigma_{U}=\hat{H}_{KS}^{\sigma}+\frac{dE^{int}}{d\tilde{\rho}^{\sigma}}=\hat{H}_{KS}^{\sigma}+V^{\sigma}_{int}[\bar{n}^{\tau\sigma}]\cdot \frac{d\bar{n}^{\tau\sigma}}{d\tilde{\rho}^{\sigma}}
\end{equation}
The $\frac{d\bar{n}^{\tau\sigma}}{d\tilde{\rho}}$ term in Eq.~\ref{energyPAW} is difficult to evaluate because the overlap matrix $O$ in Eq.$\:$\ref{eq:norm} also varies implicitly due to the change of $|\tilde{\psi}_{i\mathbf{k}}\rangle$. For simplicity, we assume that the ortho-normalization effect is fully incorporated in the electronic $V^{\sigma}_{int}[\bar{n}^{\tau\sigma}]$ term  while the change of the density matrix via $\tilde{\rho}$ is computed from the un-normalized $n^{\tau\sigma}$:
\begin{eqnarray}
V^{\sigma}_{int}[\bar{n}^{\tau\sigma}]\cdot \frac{d\bar{n}^{\tau\sigma}}{d\tilde{\rho}} &\simeq& V^{\sigma}_{int}[\bar{n}^{\tau\sigma}]\cdot \frac{dn^{\tau\sigma}}{d\tilde{\rho}}\\
&=& \sum_{ij}|\tilde{p}_i\rangle\cdot(V^{\sigma}_{int}[\bar{n}^{\tau\sigma}]\cdot\langle\phi_{i}|\hat{P}^{\tau}|\phi_{j}\rangle)\cdot\langle\tilde{p}_j|.\nonumber 
\label{eq:F_int2}
\end{eqnarray}
This interaction potential part is expanded with the basis of the projector $|p\rangle$ and it can be added to the augmentation part of $\hat{H}^{\sigma}_{KS}$ in Eq.$\:$\ref{eq:H_PAWU}. 
Therefore, the Hamiltonian $\hat{H}^\sigma_{U}$ is given by
\begin{equation}
\hat{H}^{\sigma}_{U}=-\frac{1}{2}\hat{\nabla}^2+\tilde{v}_{eff}+\sum_{i,j}|\tilde{p}_i\rangle(\tilde{D}_{ij}+D^1_{ij}-\tilde{D}^1_{ij}+\overline{V}_{ij})\langle\tilde{p}_j|
\label{eq:H_PAWU}
\end{equation}
where
\begin{equation}
\overline{V}_{ij}=V^{\sigma}_{int}[\bar{n}^{\tau\sigma}]\cdot\langle\phi_{i}|\hat{P}^{\tau}|\phi_{j}\rangle.
\end{equation}

The total energy functional within PAW  can be obtained as follows:
\begin{eqnarray}
E[\rho^\sigma,n^{\tau\sigma}]&=&\sum_{i\mathbf{k}}f_{i\mathbf{k}}\langle \tilde{\Psi}_{i\mathbf{k}}|\hat{H}_U^{\sigma}|\tilde{\Psi}_{i\mathbf{k}}\rangle+E^{PAW}_{dc}[\tilde{n},\hat{n},n^1,\tilde{n}^1]\nonumber \\
&& -\mathbf{Tr}(V^{\sigma}_{int}[\bar{n}^{\tau\sigma}]\cdot n^{\tau\sigma})+E^{int}[\bar{n}^{\tau\sigma}].
\label{eq:E_PAWU}
\end{eqnarray}
The interaction energy correction $E^{int}$ ($E^{pot}$ (Eq.$\:$\ref{eq:E_pot}) - $E^{DC}$ (Eq.$\:$\ref{eq:E_DC})) term and the interaction potential correction $V^{\sigma}_{int}$ term are computed using the orthonormalized density matrix $\overline{n}^{\tau\sigma}_{mn}$. In practice, the band energy correction term $\mathrm{Tr}(V^{\sigma}_{int}[\bar{n}^{\tau\sigma}]\cdot n^{\tau\sigma})$ is computed only updating the $V^{\sigma}_{int}$ term while the density $n^{\tau\sigma}$ is obtained from un-normalized projector functions. In this way properly orthonormalized correlated orbitals can be used with only a slight modification of the PAW formalism.

\subsection{MLWF orbitals}
\label{sec:Theory-2-2}

Here, we derive the +U formalism in the case where the correlated subspace $n^{\tau\sigma}_{mn}$ is defined by Wannier functions. We follow the  approach used in our previous analysis of the DFT+DMFT formalism\cite{Park2014235103}. In this subsection we present the formalism purely for DFT+U, as all of the comparisons between projectors and Wannier in this study will take place in the context of DFT+U. The generalization to  SDFT+U is however straightforward.

Wannier functions are discussed at length in the literature \cite{Marzari:97,Marzari:12,Wannier}. Here we make only a few remarks. First, the construction of a Wannier functions requires the specification of a hybridization window $W$, a range of energies from which the states used in the construction of the Wannier functions are defined. This energy range should encompass both the  correlated orbitals and the orbitals which directly hybridize with the correlated ones. For example, in the case of the  rare-earth nickelates, Wannier functions are constructed from an energy window ($\approx 11$ eV wide) including the  full Ni-3$d$ and O-2$p$ manifolds.  By construction the  Wannier functions provide a complete orthonormal basis for states within the energy window so it is not necessary to introduce an overlap matrix.  A continuous infinity of choices of Wannier basis exists; here we  choose a ``maximally localized'' (MLWF \cite{Marzari:97}) basis set that minimizes   the sum of Wannier function spreads ($\langle r^2 \rangle$-$\langle \mathbf{r} \rangle^2$) and also perform an additional orbital rotation as described below Eq.~\ref{HKSWannier}  We denote the resulting states as $\left|W^{\mathbf{R}_\tau}_n\right>$ where $\tau$ labels an atom within a unit cell, $\mathbf{R}$ denotes a lattice vector,  and $n$ is an orbital index.

The projection of the  DFT Hamiltonian onto the MLWF basis set is
\begin{equation}
H^{0,\mathbf{R}^\prime_{\tau^\prime}\mathbf{R}_{\tau}}_{mn}=\langle W_n^{\mathbf{R}^\prime_{\tau^\prime}}|\hat{H}_{KS}|W_m^{\mathbf{R}_{\tau}}\rangle.
\label{HKSWannier}
\end{equation}  
As discussed e.g. in Ref.~\onlinecite{Park2014235103}, for the Wannier functions pertaining to the correlated states we perform a rotation in the orbital indices to minimize the off-diagonal terms of the on-site correlated-state $\hat{H}^{0,\mathbf{R}_{corr,\tau},\mathbf{R}_{corr,\tau}}_{mn}$ in the $mn$ subspace. 

The DFT+U calculation solves the eigenvalue problem of the $\hat{H}_{KS}+\hat{V}^{\sigma}_{int}$ matrix within the hybridization window $W$. One should note that the $\hat{V}^{\sigma}_{int}$ term is spin-dependent while $\hat{H}_{KS}$ has no explicit spin dependence and has parameters determined by the total density.

The density matrix $\eta$ within the hybridization window, which includes the correlated subspace as a subset, is obtained from the eigenvalues  and eigenfunctions  of $\hat{H}_{KS}+\hat{V}^{\sigma}_{int}$ as
\begin{eqnarray}
\eta^{\tau\sigma}_{mn} &=& \sum_{l\in W,\mathbf{k}}f(\epsilon^{\sigma}_{l\mathbf{k}})\langle\psi_{l\mathbf{k}}|W_m^{\mathbf{R}_{\tau}}\rangle\langle W_n^{\mathbf{R}_{\tau}}|\psi_{l\mathbf{k}}\rangle\label{eq:n_wan}.
\end{eqnarray}
The density matrix $n^{\tau\sigma}_{mn}$ in the correlated subspace is a sub-block of $\eta^{\tau\sigma}_{mn}$. Our basis choice in the $nm$ space means that the  off-diagonal terms are negligible;   $n^{\tau\sigma}_{mn}\approx \delta_{nm}$.

The band energy correction $E^{\Delta KS}$ is then given by 
\begin{equation}
E^{\Delta KS} =\mathbf{Tr}(\hat{H}_{KS}\cdot \eta)-\mathbf{Tr}(\hat{H}_{KS}\cdot \eta^{0}),
\label{eq:band_Wan}
\end{equation}
where the $\eta^{0,\tau}_{mn}$ is computed via Eq.~\ref{eq:n_wan} but using the eigenfunctions and eigenvalues of  $\hat{H}_{KS}$ rather than $H_{KS}+V^{\sigma}_{int}$.

The interaction energy can be defined using Eq.$\:$\ref{eq:E_pot} with the calculated density matrix (Eq.$\:$\ref{eq:n_wan}). In our application to the nickelates  we construct this  term using the Slater-Kanamori Hamiltonian as defined in our previous paper~\cite{Park2014235103} and for ease of reference present the results in the same notation. In Ref. \onlinecite{Park2014235103} the on-site intra-orbital interaction is given as $u$, the Hund's coupling is $j$  the inter-orbital interaction is  $u-2j$ and the exchange and pair-hopping terms do not contribute in the Hartree-Fock approximation used here.  The parameters are related to the $U$ and $J$ defined elsewhere in the paper by $u=U+(8/7)J$ and $j=(5/7)J$. For the specific application to the nickelates we took 
$u$=6.14eV and $j$=0.71eV (corresponding to $U$=5eV and $J$=1eV).

The interaction potential energy is then
\begin{eqnarray}
E^{pot} &=& u\sum_{m,\tau} n^{\tau\uparrow}_{m}n^{\tau\downarrow}_{m} + (u-2j)\sum_{m\neq m',\tau} n^{\tau\uparrow}_{m}n^{\tau\downarrow}_{m'} \nonumber \\
&&+ (u-3j)\sum_{m>m',\tau\sigma} n^{\tau\sigma}_{m}n^{\tau\sigma}_{m'}
\label{eq:HF2}
\end{eqnarray}
while the double counting energy is 
\begin{eqnarray}
E^{DC} & = & \frac{u}{2}N_d(N_d-1)-\frac{5j}{4}N_d(N_d-2),
\label{eq:DC2}
\end{eqnarray}
where here $N_d$ is the total occupancy of the $d$ levels on a Ni ion.

\section{ The Force functional}
\label{sec:Theory-3}

The force functional is defined in terms of the derivatives of the energy functional with respect to atomic positions. Here we present the force functional corresponding to the DFT+U version of the energy functional in Eq.~\ref{eq:dftu}; the SDFT+U forces can be derived similarly.  The specifics depend on the formalism. As before, the PAW formalism as implemented in VASP is utilized and we present forces for the projector, while we outline the differences for the case of a  Wannier based correlated subspace. 

The DFT+U force functional consists of two parts, namely the same functional form used in DFT except that the DFT Fermi function is replaced by the DFT+U density matrix utilizing DFT+U eigenvalues and eigenfunctions and an additional force term derived from the $E^{int}$ energy term. The computation of the forces  requires consideration of the  derivatives of the correlated orbital density matrix with respect to $\mathbf{R}$, which in turn arise from the changes of the correlated orbital wave functions as the atomic positions change. The derivative of the projector orbital is already computed within DFT force formalism, therefore one can adopt the the same calculation already implemented in VASP. In the case of Wannier functions, additional implementation is needed to compute the derivative of Wannier function $|W_m^{\mathbf{R}}\rangle$ with respect to $\mathbf{R}$.

The force functional following from the projector correlated orbital set can be derived from the energy functional in  Eq.$\:$\ref{eq:E_PAWU}; taking derivatives with respect to $\mathbf{R}$ produces the same functional form as the PAW force functional (see e.g. Ref.~\onlinecite{Kresse19991758} for details) except that the eigenvalues $\epsilon_n$ and eigenfunctions $\tilde{\psi}_n$ are obtained by solving $\hat{H}_{U}^{\sigma}$. The force has terms due to the change of the pseudized core charge density $\tilde{n}_{Zc}$ via the explicit movement of the ionic positions, the change of the compensation charge $\hat{n}$ itself, and the change of the projector functions $|\tilde{p}\rangle$ as the ions are moved:
\begin{eqnarray}
F^{DFT+U} &=&  -\sum_{n} f_n \Big\langle\tilde{\psi}_n\Big|\frac{\partial (\hat{H}^{\sigma}_U-\epsilon_n(1+\sum_{ij}|\tilde{p}_i\rangle q_{ij}\langle\tilde{p}_j|))}{\partial \mathbf{R}}\Big| \tilde{\psi}_n \Big\rangle \nonumber \\ 
&=& F^1\Big[\frac{\partial\tilde{n}_{Zc}}{\partial\mathbf{R}}\Big]+F^2\Big[\frac{\partial\hat{n}}{\partial\mathbf{R}}\Big]+F^3\Big[\frac{\partial|\tilde{p}\rangle\langle\tilde{p}|}{\partial\mathbf{R}}\Big]
\label{eq:F_PAWU}
\end{eqnarray}
where
\begin{eqnarray}
F^3\Big[\frac{\partial|\tilde{p}\rangle\langle\tilde{p}|}{\partial\mathbf{R}}\Big]=& -& \sum_{n,ij} (\hat{D}_{ij}+D^1_{ij}-\tilde{D}^1_{ij}+\overline{V}_{ij}-\epsilon_nq_{ij})
\nonumber \\
&\times& f_n \langle\tilde{\Psi}_n|\frac{\partial |\tilde{p}_i\rangle\langle\tilde{p}_j|}{\partial\mathbf{R}}|\tilde{\Psi}_n\rangle. 
\label{eq:F_DFT2}
\end{eqnarray}
$q_{ij}$ is the correction to the overlap matrix given by $\langle\phi_i|\phi_j\rangle-\langle\tilde{\phi}_i|\tilde{\phi}_j\rangle$.
The explicit expressions of $F^1$ and $F^2$ are the same as DFT forces given in Ref.~\onlinecite{Kresse19991758}.
The implicit changes of the Hamiltonian $H^{\sigma}_U$ via the density $\tilde{n}$, $n^1$, $\tilde{n}^1$, and $\hat{n}$ are always cancelled out exactly against the change of $E^{PAW}_{dc}$ terms in Eq.$\:$\ref{eq:E_PAWU}.

The evaluation of a force term from $E^{int}$ requires the derivative of the correlated orbital density matrix, i.e. $\frac{dn^{\tau\sigma}}{d\mathbf{R}}$ and the result depends on the choice of correlated orbital sets.
Within the projector scheme, the force term arising from the implicit change of interaction energy correction $E^{int}$ (Eq.$\:$\ref{eq:E_PAWU}) via the ortho-normalized density matrix $\bar{n}^{\tau\sigma}$ is given by
\begin{equation}
F^{int}=\frac{dE^{int}[\bar{n}^{\tau\sigma}]}{d\bar{n}^{\tau\sigma}}\cdot \frac{d\bar{n}^{\tau\sigma}}{d\mathbf{R}}=V^{\sigma}_{int}[\bar{n}^{\tau\sigma}]\cdot \frac{d\bar{n}^{\tau\sigma}}{d\mathbf{R}}.
\label{eq:F_int}
\end{equation}
Here, the calculation of $\frac{d\bar{n}^{\tau\sigma}}{d\mathbf{R}}$ term is complicated by the $\mathbf{R}$-dependence of  the overlap matrix $O$ (Eq.$\:$\ref{eq:norm}). In practical applications the derivative of the density matrix with respect to the ionic position $\mathbf{R}$ is thus approximated to the change of the un-normalized $n^{\tau\sigma}$ via the derivative of the projector function which is already present in PAW force (Eq.$\:$\ref{eq:F_DFT2}):
\begin{equation}
F^{int}\simeq V^{\sigma}_{int}[\bar{n}^{\tau\sigma}]\cdot \frac{dn^{\tau\sigma}}{d\mathbf{R}}.
\label{eq:F_int2}
\end{equation}
Taking a derivative of the $-\mathbf{Tr}(V^{int}[\bar{n}^{\tau\sigma}]\cdot n^{\tau\sigma})$ term in Eq.$\:$\ref{eq:E_PAWU} with respect to $\mathbf{R}$ leads to a term $-V^{int}[\bar{n}^{\tau\sigma}]\cdot \frac{dn^{\tau\sigma}}{d\mathbf{R}}$ which cancels out the term in Eq.$\:$\ref{eq:F_int2} and a term $-\frac{dV^{int}[\bar{n}^{\tau\sigma}]}{d\mathbf{R}}\cdot n^{\tau\sigma}$ which cancels out the implicit change of $V^{int}[\bar{n}^{\tau\sigma}]$ term in $\hat{H}^{\sigma}_U$.

The force functional using Wannier functions can be derived in a similar way as the projector functions except that the $\frac{dn^{\tau\sigma}}{d\mathbf{R}}$ term needs to be computed explicitly in terms of  the change of Wannier functions with $\mathbf{R}$. However, we have not yet implemented 
this. Instead, we have determined the minimum energy structure in the phase space of pressure and bond length difference as defined in our previous paper\cite{Park2014235103}.

\section{Application to structural properties of rare-earth nickelates}
\label{sec:Theory-4}

\subsection{Overview}
In following sections we investigate the general issues of  interest in this paper, namely different correlated orbital sets (Projector vs Wannier) and background electronic structure methods  (DFT+U vs SDFT+U), in the specific context of the structural properties of the rare-earth nickelates, $R$NiO$_3$.  In these materials the basic structural motif is the NiO$_6$ octahedron. At some values of temperature, pressure and $R$, the materials exhibit  a uniform phase in which all octahedra have approximately the same mean Ni-O bond length. In other parameter regimes the materials exhibit a two-sublattice disproportionated phase in which the octahedra on one sublattice have a mean Ni-O bond length $\sim 0.1$\AA\  shorter than the octahedra on the other sublattice.  The materials are important for the present study because this basic structural property  is closely linked to a fundamental  electronic property,  namely whether the material is a metal or a correlation-driven site-selective Mott insulator \cite{Park:12,Alonso:99}. This linkage means that obtaining a correct description of the structural properties poses a critical test for the electronic structure methods. 

Our previous studies\cite{Park2014235103,Park:14} showed that (S)DFT+U does not provide a quantitatively accurate description of the experimental structural and metal-insulator transition phase diagram of  $R$NiO$_3$  series. A particular difficulty is the prediction that the  ambient-pressure ground state of LaNiO$_3$ is bond-disproportionated and insulating when the actual material has a non-disproportionated $R\bar{3}c$ structure and is metallic. A closely related deficiency of the DFT+U method is an overestimation of  the critical pressure of the metal-insulator  transition for materials where the ambient pressure ground state is insulating.  DFT+DMFT methods produced much better results. It is also the case that DFT+U (and DFT+DMFT) wrongly predict that the ground state is ferromagnetic.

However the trends found in the DFT+U calculations were found to track the trends found in the DFT+DMFT calculations. For example  the DFT+U $T\rightarrow 0$ structural phase boundary  in the pressure-tolerance factor plane was offset by a certain pressume from the DFT+DMFT phase boundary, indicating that   the difficulty is simply that the DFT+U methods overestimate (to a considerable degree) the stability of the insulating state.  Thus since the aim of the current investigation is understand how  different formulations of the theory affect basic issues of structure and energetics, rather than to accurately model material properties, we can use the +U approximation as a flexible and inexpensive computational laboratory with the expectation that the similarity of trends noted above suggests that the general findings will be applicable to DFT+DMFT as well.

We present results computed as described in Appendix~\ref{Appendix:sec:Theory-4}.  
In the current paper, we use the Vienna Ab-initio Simulation Package (VASP)~\cite{Kresse199611169,Kresse19991758}
which adopts the PAW formalism.
For the exchange-correlation DFT functional, we use a generalized gradient approximation (GGA) 
with the Perdue-Burke-Ernzerhof (PBE) functional\cite{Perdew19963865} and also adopt a local density approximation (LDA) if necessary.
We take the correlated orbitals to be atomic-like Ni-centered $d$ orbitals using the projector method and assume the additional interactions have the form given in Eq.$\:$\ref{eq:E_pot}. Unless otherwise specified, we use $U$=5eV and $J$=1eV for all computations.


We present results for three members of the material family: LuNiO$_3$ (strong insulator at ambient pressure), NdNiO$_3$ (insulating but near the phase boundary at ambient pressure), and LaNiO$_3$ (metallic at ambient pressure).  
In the rest of this section we introduce the materials and compare the DFT+U and SDFT+U results using both the GGA and LDA functionals for bond lengths and energetics. In the following section we  discuss the projector vs Wannier issue and in a third section present implications for computed phase diagrams. 


\subsection{Bond disproportionation vs volume}

\begin{figure}[!htbp]
\includegraphics[scale=0.45]{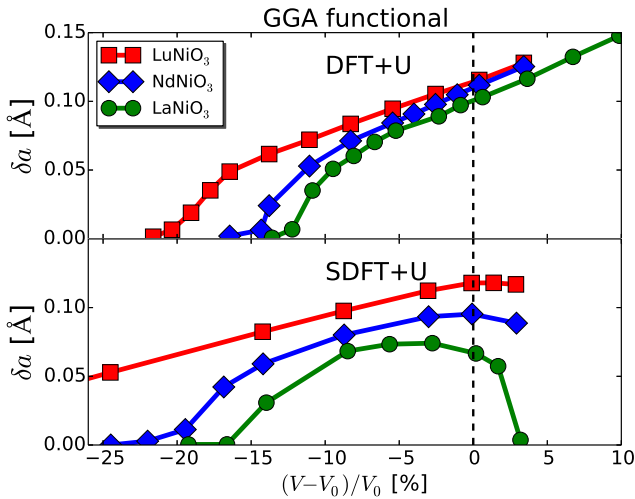}
\caption{(Color online) 
Average Ni-O bond-length difference $\delta a$ as a function of volume computed using the GGA functional 
with DFT+U (top panel) and SDFT+U (lower panel) for LuNiO$_3$ (red square), NdNiO$_3$ (blue diamond), and LaNiO$_3$ (green circle). 
$U$=5eV and $J$=1eV are used.  
The same orthonormalized projector method is used to construct the correlated subspace in all cases.  
$V_0$ is the zero-pressure volume computed for the given compound by the given method. 
\label{fig:bond_sp}}
\end{figure}

We used the VASP implementation of (S)DFT+U using the GGA functional to perform full structural relaxations of the three compounds 
from the low-symmetry bond-disproportionated structure
with the unit cell volume constrained to take particular values (see Fig.$\:$\ref{fig:bond_sp}).
The  correlated orbital basis set was treated  using the same orthonormalized projector for all calculations (note that the orthonormalization of the basis set required that we modify the VASP energy and force formulas according to Eq.$\:$\ref{eq:norm}). In many cases a disproportionated structure with two inequivalent NiO$_6$ octahedra was found; for these cases   we computed the difference $\delta a$ in mean Ni-O bond length between Ni sites on different sublattices. Generically if $\delta a \neq 0$ the band structure exhibits a gap at the fermi level (for very small disproportionation amplitudes the gapping may not be complete), but for simplicity of presentation we do not consider the electronic structure here.   

Results are shown in Fig.$\:$\ref{fig:bond_sp}. 
At positive compression ($V-V_0<0$) the differences between SDFT+U and DFT+U are quantitative, but large. 
We see that consistently across the material family SDFT+U predicts a higher  $\delta a$ at given compression and similarly predicts that a higher critical compression is  needed to drive  the structural transition ($\delta a\rightarrow 0$) than does  DFT+U.  However, at negative compression ($V-V_0>0$) the difference is qualitative:  DFT+U predicts a monotonic increase of $\delta a$ values as $(V-V_0)/V_0$ increases while SDFT+U calculations indicate a reduction of $\delta a$ as the cell volume is increased and ultimately a reentrant structural transition (seen in the data for LaNiO$_3$ and expected for the other materials from the downward curvature). 


\begin{figure}[!htbp]
\includegraphics[scale=0.45]{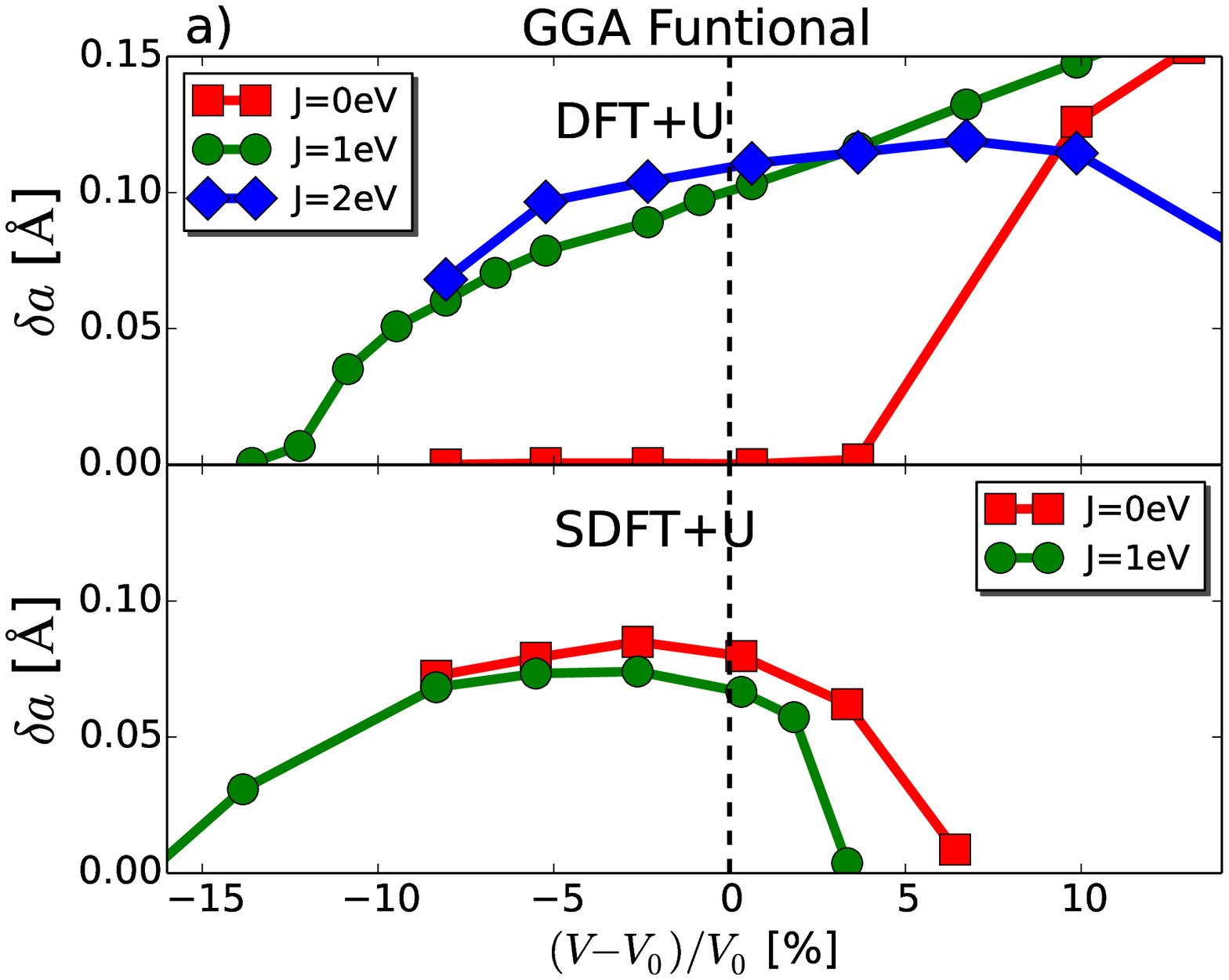}
\includegraphics[scale=0.45]{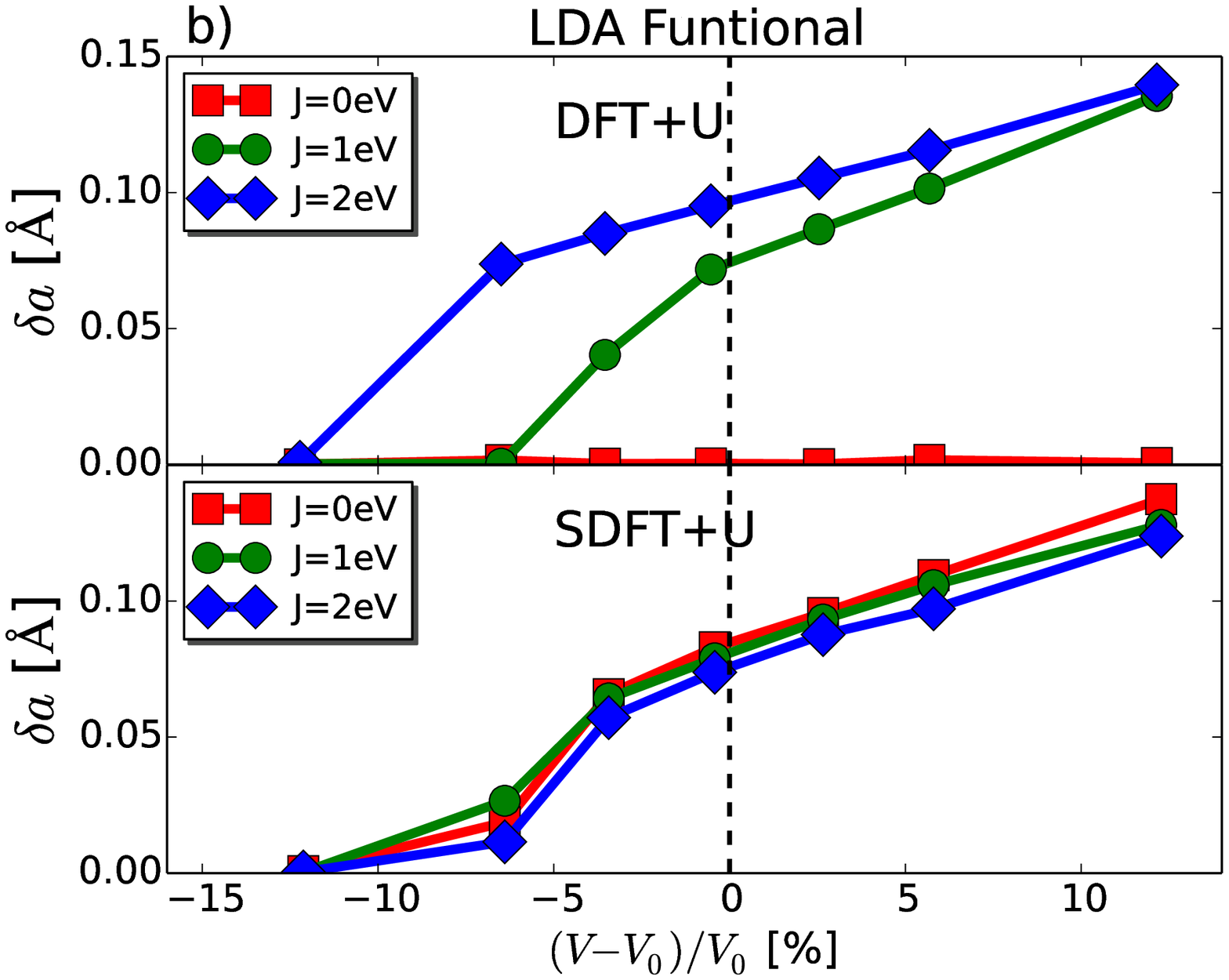}
 \caption{(Color online) 
Bond disproportionation $\delta a$  plotted against relative volume change $(V-V_0)/V_0$ computed for LaNiO$_3$ for different Hunds coupling $J$ at $U$=5eV using both (a) GGA and (b) LDA functionals and different methods indicated in the legends (DFT+U vs SDFT+U). The same orthonormalized projector method is used to construct the correlated subspace in all cases.  $V_0$ is the zero-pressure volume computed using the given method and $J$=1eV. The change of $V_0$ at different $J$ is negligible for SDFT+U and rather notable for DFT+U (see Fig.$\:$\ref{fig:E_tot}) but it does not affect any results discussed here. 
 \label{fig:La_bond}}
\end{figure}

Fig.$\:$\ref{fig:La_bond} shows the effect of varying the Hunds coupling $J$ on the computed bond disproportionation for LaNiO$_3$.  In the DFT+U calculations, increasing $J$ from $0$ to $1$ eV has a dramatic effect, while a further increase to $2$ eV  has a weaker effect, suggesting a saturation as J is increased. On the other hand, the SDFT+U results show almost no J-dependence,  
indicating that  the spin dependent exchange potential in SDFT already effectively includes a large on-site $J$ and suggesting  that $J$ is not needed when performing SDFT+U calculations. This could be problematic for SDFT+DMFT calculations, where the dynamical effect of $J$ are typically  important.

The upper portions of the two panels of Fig. ~\ref{fig:La_bond}  further show that within DFT+U the choice of DFT method (LDA vs GGA) produces quantitative but not qualitative differences over the volume range investigated, with in particular the LDA+U exhibiting a smaller $\delta a$ at given $J$ and volume, consistent with the known tendency of the PBE GGA functional used here to overestimate magnetism~\cite{Perdew2008136406}.
Alternatively, in SDFT+U the choice of DFT method produces a qualitative difference, with the SGGA+U method indicating reentrance of the non-disproportionated phase at a small positive relative volume while no indication of reentrance is found in the LSDA+U calculations. Interestingly, The $J$=2eV GGA+U calculations also suggest that reentrance of the undistorted phase would occur at larger relative volumes, consistent with the notion that the SDFT methods imply a large (perhaps excessively  large) J already at the SDFT level. Taken together these results also suggest that the mathematical origin of the reentrant transition is a (presumably unphysical) effect of large J. The disproportionated phase may be understood as a hybridization density wave corresponding to relatively strong Ni-O bonding at the short-$\delta-a$ site ~\cite{Park:14} so although the precise connection is not clear at this point we may speculate that the reentrance is related to unphysically large spin-dependence of level shifts of the Ni-$d$ relative to O-$p$ states, weakening the Ni-O singlet bond that produces the distortion.

\begin{figure}[!htbp]
\includegraphics[scale=0.45]{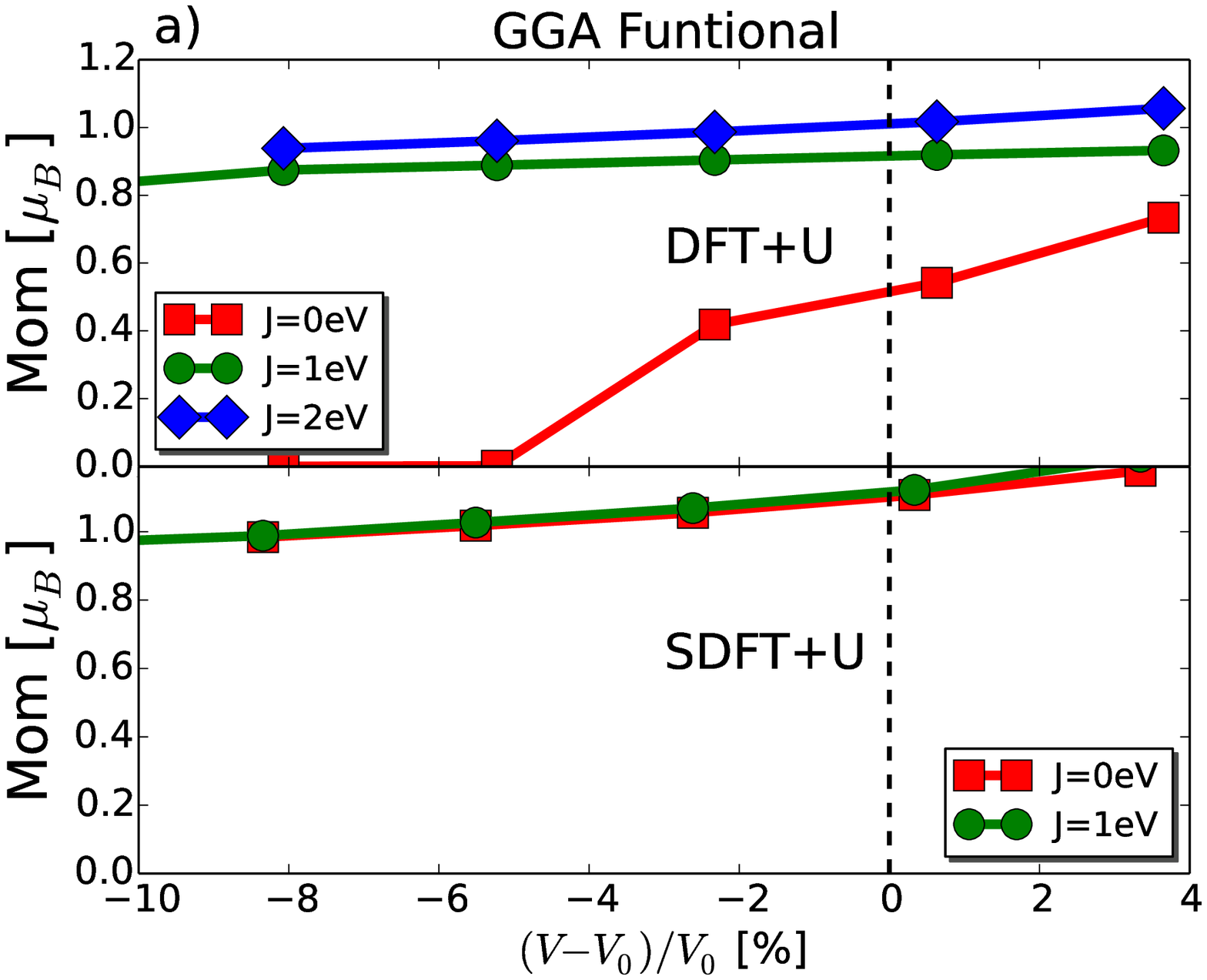}
\includegraphics[scale=0.45]{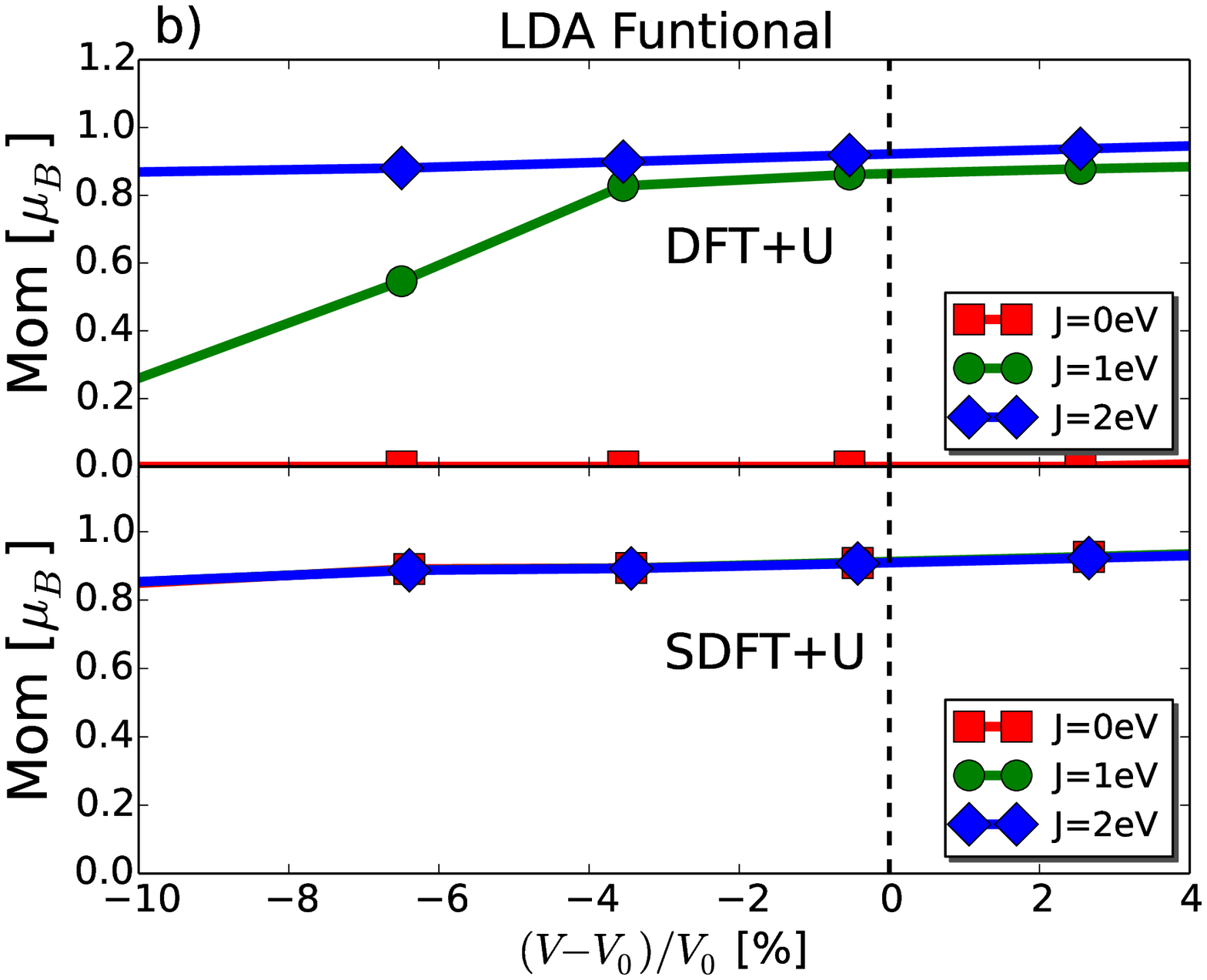}
 \caption{(Color online) 
The magnetic moments per Ni atom in LaNiO$_3$ obtained using different $J$ as a function of pressure using both (a) GGA and (b) LDA functionals and different methods indicated in the legends (DFT+U vs SDFT+U). The same orthonormalized projector method is used to construct the correlated subspace in all cases.  $V_0$ is the zero-pressure volume computed using the given method and $J$=1eV. 
\label{fig:mom}}
\end{figure}

The interplay between the DFT functional (LDA vs GGA), the value of Hund's $J$ and the physics of the disproportionation instability are also evident in the study of the magnetic moments presented in Fig.$\:$\ref{fig:mom}. The DFT+U results reveal the expected dependence of magnetic moment on $J$, with magnetic moment increasing with $J$ with the dependence becoming weaker as the  saturation  value  $M=1~\mu_B$ is reached and the critical volume for the magnetic transition also being $J$-dependent. The difference between LDA+U and GGA +U results reflects the stronger tendency toward magnetism  characteristic of the PBE-GGA functional. In effect, PBE-GGA already contains a certain degree of local exchange.  In contrast, SDFT+U calculations in both GGA and LDA produce large moments at all volumes, and with negligible $J$  dependence. 

\section{Energetics}

\begin{figure}[!htbp]
\includegraphics[scale=0.45]{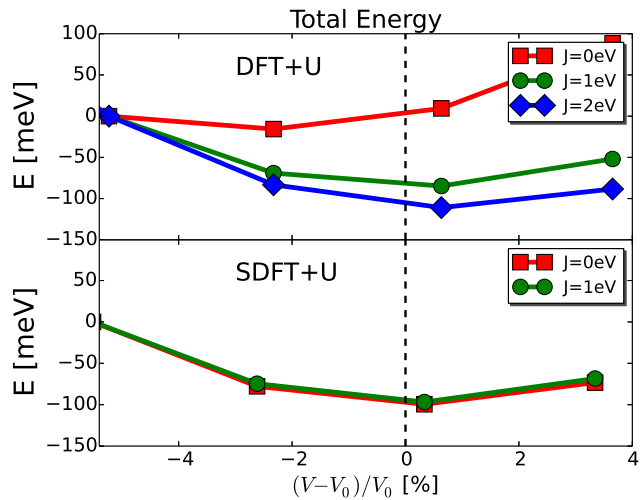}
\caption{(Color online) 
The total energy of LaNiO$_3$ as a function of relative volume difference computed using GGA+U (top) and SGGA+U (bottom) with an orthonormalized projector definition of the correlated orbitals at  $J$ values indicated. At each volume and for each method the structure is relaxed and the energy of the relaxed structure is presented. The zero of energy is chosen at a compression of  $5\%$ in all curves. 
 \label{fig:E_tot}}
\end{figure}

In this section, we compare the DFT and SDFT predictions for energies.  We restrict attention to the GGA and SGGA density functionals, define the correlated states via orthonormalized projectors, and focus on LaNiO$_3$. For each relative volume, the structure was relaxed and then the energy was evaluated. Fig$\:$\ref{fig:E_tot} displays the dependence of the total energy on the normalized volume difference for different $J$ values. 

As found in the previous section's analysis of the disproportionation amplitude and magnetic moment, substantial differences between DFT+U and SDFT+U are found. The DFT+U energy curve depends substantially on $J$, changing rapidly as $J$ is increased from zero and saturating as $J$ becomes large. Remarkably even the equilibrium volume is $J$-dependent. The SDFT+U energy has negligible $J$  dependence and is similar to the $J$=2eV DFT+U result again suggesting that the SDFT exchange correlation functional in effect contains a $J$ which (for the PBE-GGA case studied here) is substantially larger than the $J\sim$ 1eV values believed to be physically reasonable. 

\begin{figure}[!htbp]
\includegraphics[width=\linewidth]{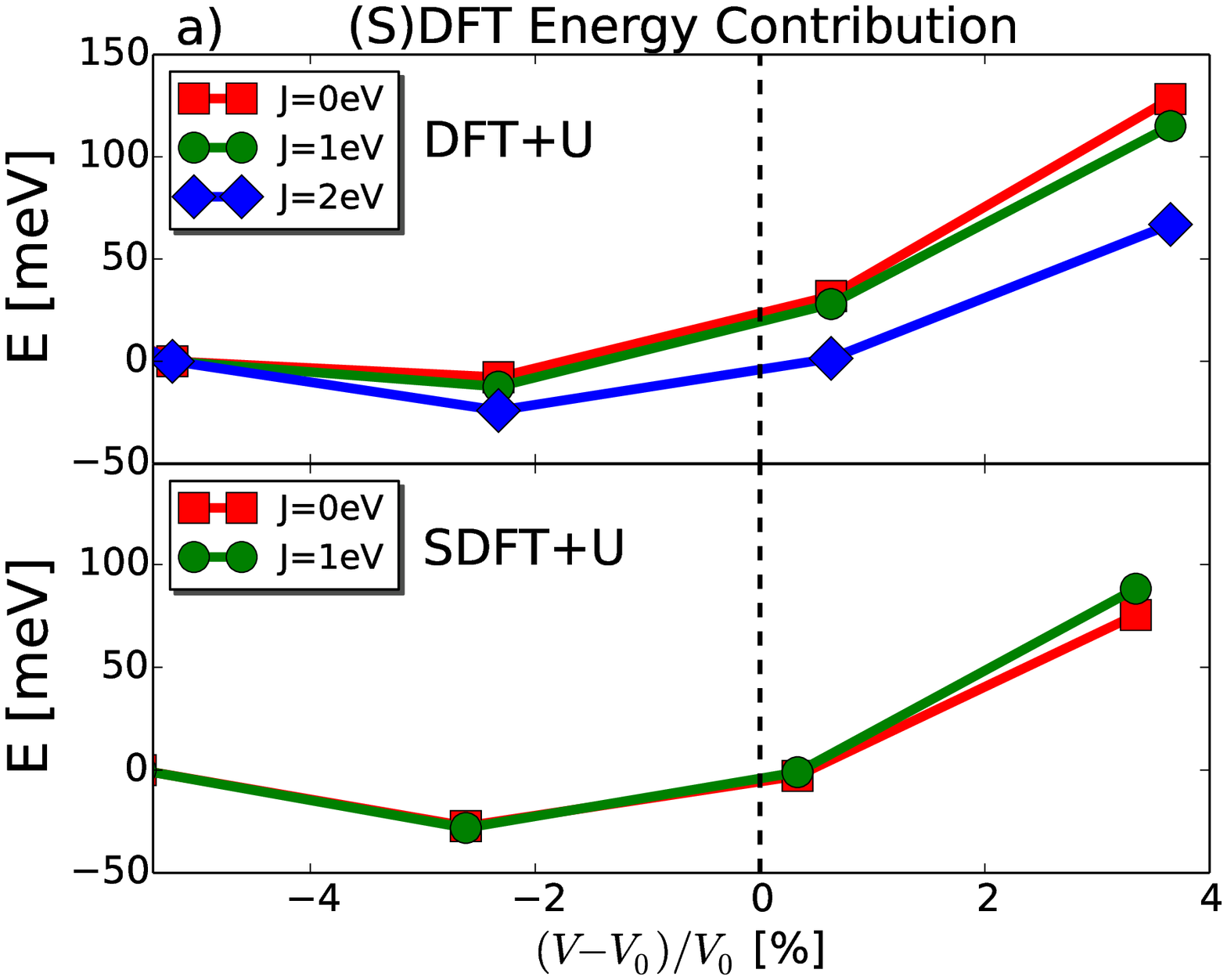}
\includegraphics[width=\linewidth]{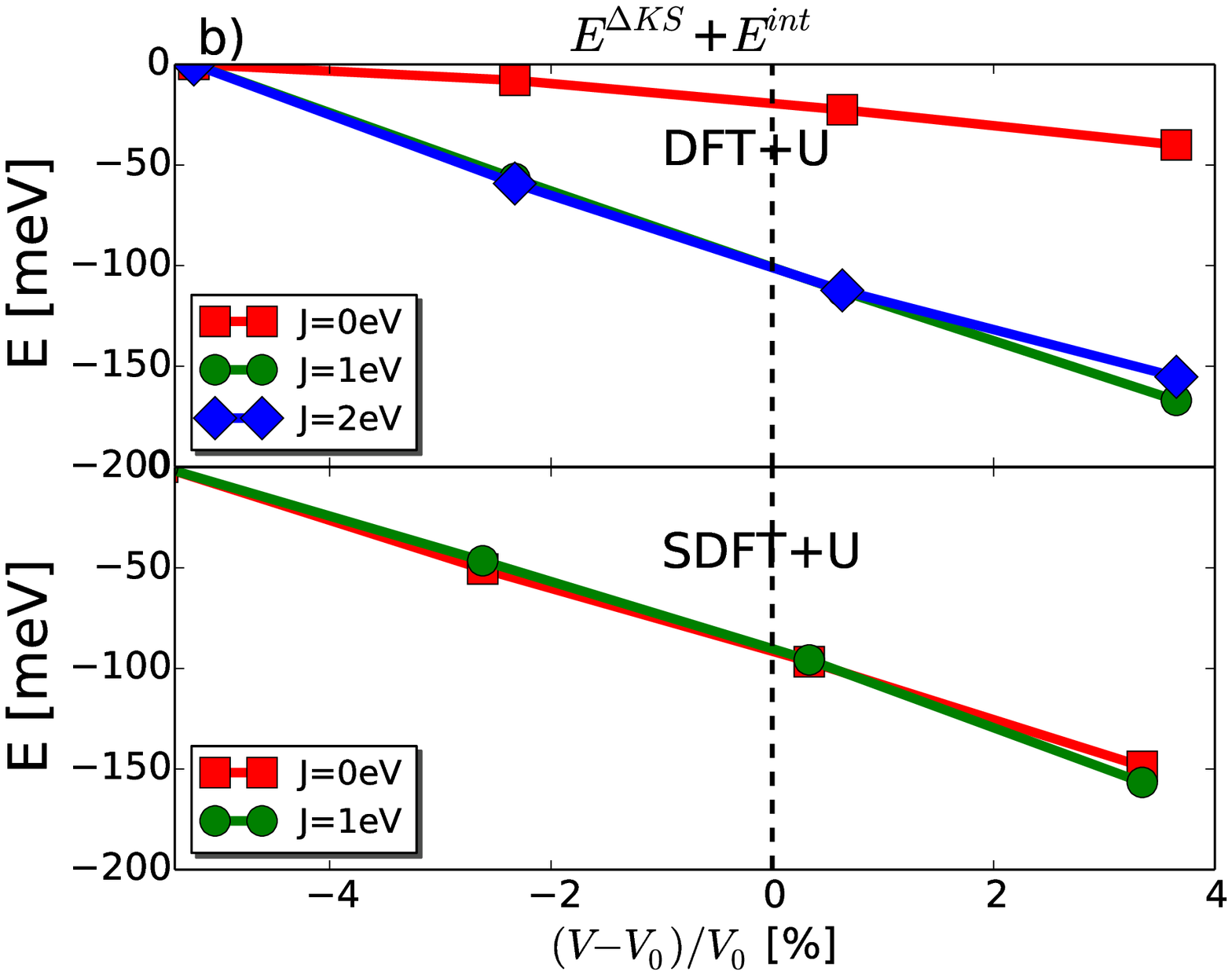}
 \caption{(Color online) 
Contributions to DFT+U energy functional (cf Eq.~\ref{eq:Energy}) computed for  LaNiO$_3$ as a function of relative volume.  
a) The DFT contribution  and b) the correlation correction CC contribution ($E^{\Delta KS}+E^{int}$) are decomposed from Fig.$\:$\ref{fig:E_tot} using the same relaxed structure at each volume and compared for different methods (DFT+U (the top panel) vs SDFT+U (the bottom panel)) and different $J$ values. 
\label{fig:E_DFT}}
\end{figure}

Fig.$\:$\ref{fig:E_DFT} presents a decomposition of the energy into the DFT contribution ($E^{DFT}$,  panel a) and the correlation correction (CC) contribution ($E^{\Delta KS}+E^{int}$, lower panel), as defined in Eq.~\ref{eq:Energy}, for DFT+U (upper half of each panel) and SDFT+U (lower half of each panel). The DFT term $E^{DFT}$ contains the structural contribution while the CC term expresses the correlation physics. $E^{DFT}$ is not monotonic in unit cell volume, expressing the basic physics of chemical bonding. Alternatively, the CC contribution  decreases monotonically as the volume is increased for both DFT+U and SDFT+U and for all $J$ values, expressing the enhancement of correlation occurring when hybridization is decreased and showing that the equilibrium volume predicted by the correlated calculation is larger than that predicted from the DFT contribution alone.

Fig.$\:$\ref{fig:E_DFT} indicates that for the SDFT+U method  neither $E^{DFT}$ nor the CC term has
significant $J$ dependence because the SDFT method already includes a large
local exchange contribution. Alternatively, in the DFT+U calculation both
terms have some $J$ dependence. The DFT+U CC term changes
dramatically as $J$ is increased from $0$ to $1eV$, accounting for the
noticeable change of DFT+U energetics from $J$=0 to 1eV in
Fig.$\:$\ref{fig:E_tot}  but does not change much as $J$ is further increased
because the moment is saturated  (see Fig.$\:$\ref{fig:mom}).  It is also
interesting to note that the DFT+U CC energy at $J\gtrsim 1eV$ is comparable to
the SDFT+U CC energy at $J=0$, further confirming the large value of $J$ implicit
in the SDFT method. In the DFT+U method, some dependence of $E^{DFT}$ on $J$
occurs as $J$ is increased from $1eV$ to $2eV$, and it is interesting to note
that this is the $J$ range where suggestions of reentrance are visible in the
DFT+U calculation. This behavior again indicates that the unusual reentrant
behavior is related to a rearrangement of the band-structure by an unphysically
large Hunds coupling.

\begin{figure}[!htbp]
\includegraphics[scale=0.45]{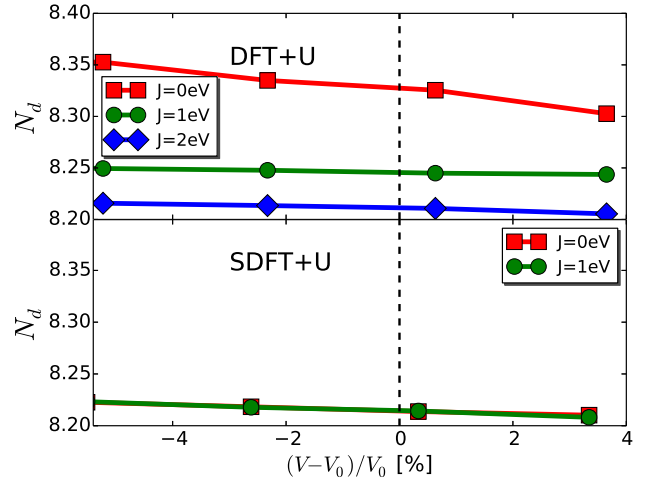}
 \caption{(Color online) 
The  occupancy of the correlated orbitals expressed as the number of $d$ electrons   $N_d$  per Ni atom computed for  LaNiO$_3$ at $J$ values indicated for GGA+U (the top panel) and SGGA+U (the bottom panel).
\label{fig:Nd}}
\end{figure}

As extensively discussed elsewhere~\cite{Wang:12,Dang:13}, the occupancy $N_d$ of the correlated orbitals provides useful insights into the physics of strongly interacting electron systems. Fig.$\:$\ref{fig:Nd} displays the $d$ occupancies computed for LaNiO$_3$ for the parameters whose energies are shown in Fig.$\:$\ref{fig:E_DFT}. Within DFT+U, increasing the Hunds coupling $J$ leads to a decrease in $N_d$, a signal of stronger correlation arising from effectively smaller $p$-$d$ hybridization. In SDFT+U the correlations in this sense are already stronger at the SDFT level (ie. SDFT+U yields smaller $N_d$), and adding additional $J$ does not change the situation. 

\section{Choice of Correlated Orbital}

\begin{figure}[!htbp]
\includegraphics[width=\linewidth]{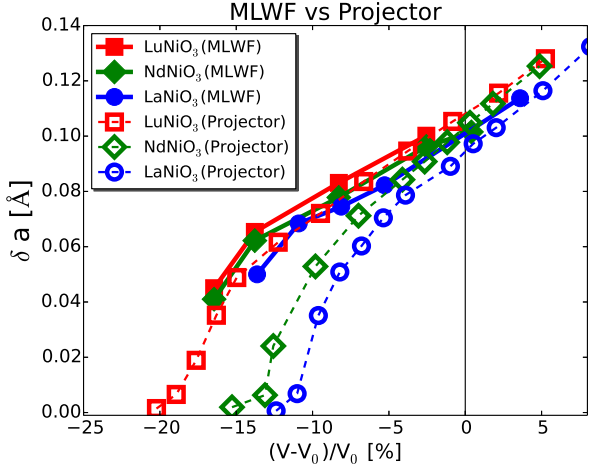}
 \caption{(Color online) 
The average Ni-O bond length difference $\delta a$ as a function of reduced volume computed for materials indicated using DFT+$U$ as implemented with the energy functional of Eq.$\:$\ref{eq:Energy}. Both maximally localized Wannier functions (filled symbols, solid lines)  and ortho-normalized projectors (open symbols, dashed lines)) are compared. Interaction parameters of $U$=5eV and $J$=1eV are used for the projector-based calcualtions while the equivalent values  $u$=6.14eV and $j$=0.71eV (Slater-Kanamori parameterization) are correspondingly used for the Wannier construction.
\label{fig:bond_basis}}
\end{figure}

In this section we study the effect of the choice of correlated orbital on the calculated results. The DFT+U method is used; SDFT+U is not considered in this section.  Fig$\:$\ref{fig:bond_basis} presents the  bond disproportionation amplitude  $\delta a$ versus reduced volume for different rare earth nickelates using either MLWF  (filled symbols, solid lines) or  ortho-normalized projectors (open symbols, dashed lines) for the correlated subspace.  The qualitative trends of $\delta a$ as a function of reduced volume  are similar for both correlated orbitals. Substantial differences appear only for very large compression, where the Wannier approach enhances  the tendency to the bond disproportionated states, though only for  NdNiO$_3$ and LaNiO$_3$. The origin of the difference is not clear at present, but may have to do with the fact that size of the Wannier function varies as the volume and the bond length difference change, while the projectors are defined using a fixed radius.

\begin{figure}[!htbp]
\includegraphics[scale=0.45]{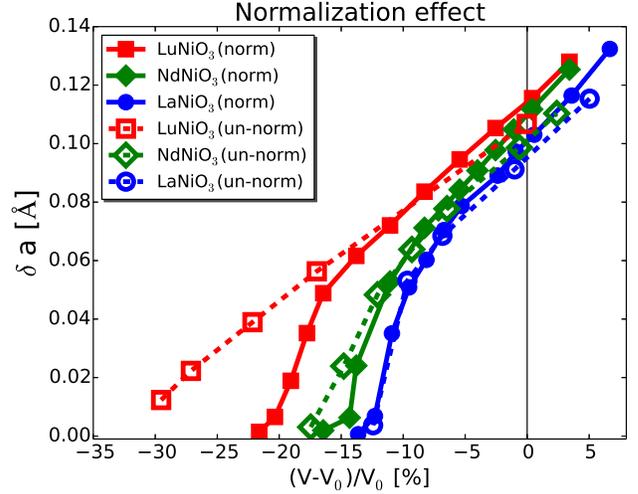}
 \caption{(Color online) 
Average Ni-O bond length difference $\delta a$ graph as a function of volumes obtained using DFT+$U$.
The normalization effect of the correlated orbital is investigated by comparing both the ortho-normalized projector correlated orbital (filled symbols, solid lines)
and the un-normalized projector (open symbols, dashed lines).
LuNiO$_3$ (red square), NdNiO$_3$ (green diamond), and LaNiO$_3$ (blue circle) results are displayed for comparison.
 \label{fig:bond_norm}}
\end{figure}

We now turn to the effect of orthonormalization, comparing in Fig.$\:$\ref{fig:bond_norm}  structural relaxation calculations performed using the orthonormalized projector orbital to calculations and performed using the unnormalized projector implemented in VASP.  Normalization has a particularly important effect in the  small volume region of LuNiO$_3$, where  the critical pressure calculated using the unnormalized projector is overestimated and probably incorrect. However, normalization has no noticeable effect on the structural relaxation of LaNiO$_3$. As the volume is expanded the consequences of normalization are seen to be minor. This discrepancy in the small volume region may arise from an overestimate of the spectral weight inside the Ni atomic sphere, leading to a mis-estimate of the correlation energy.

\section{The phase diagram of $R$NiO$_3$ and the equilibrium volume $V_0$}
\label{sec:Results}

In this section, we compute the structural phase digrams and equilibrium volume $V_0$ of the rare-earth nickelates $R$NiO$_3$  obtained using DFT+U and SDFT+U as functions of the reduced volume for different $R$ ions. MLWF and project definitions of the correlated orbitals are also explored.

\begin{figure}[!htbp]
\includegraphics[scale=0.45]{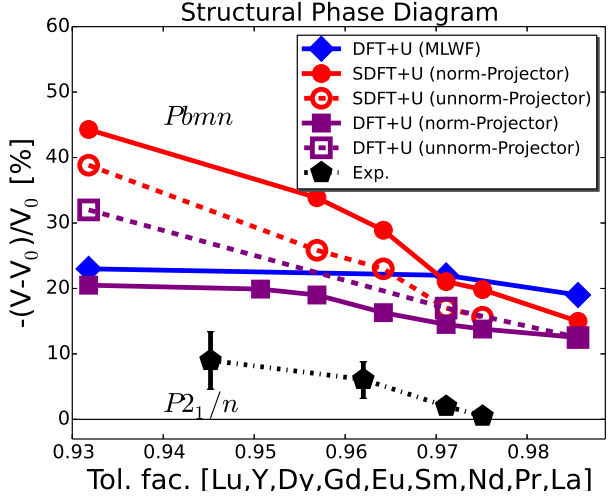}
\caption{(Color online)
Structural phase diagram of the rare-earth nickelates as a function of the volume compression and rare earth ion,  computed using DFT+U (square dots)  and SDFT+U (circle dots). The un-normalized projector  (open dots) and  ortho-normalized projector (filled dots) are compared for both methods. The MLWF correlated orbital result (diamond dot) is also shown for DFT+U, yielding some  similarity to the  phase boundary to the ortho-normalized projector. The experimental data (black dash-dot lines) are also given for comparison\cite{Cheng:10} (see also Ref.~\onlinecite{Park2014235103,Park:14}).
\label{fig:phased_sp}}
\end{figure}

Fig.$\:$\ref{fig:phased_sp} displays the structural transition  phase diagram of rare-earth nickelates $R$NiO$_3$ in the plane of reduced volume $(V-V_0)/V_0$) and choice $R$ of rare earth ion computed using the different beyond DFT methods discussed in this paper.  The structural transition is between the  $P2_1/n$ structure ($\delta a>$0) and $Pbnm$ structure ($\delta a$=0) for all $R$NiO$_3$ except rhombohedral LaNiO$_3$. For LaNiO$_3$, the transition associated with the bond-disproportionation separates the $R3$ structure ($\delta a>$0) and the $R\bar{3}c$ structure ($\delta a$=0).

All DFT+U and SDFT+U results produce critical compression for  the  transition which is too large relative to  experiment. (DFT+DMFT produces results in much better agreement with experiment\cite{Park:14,Park2014235103}). The  SDFT+U method (red circular dots) exhibits more rapid variation of the critical compression with  change of $R$ ions than does DFT+U (purple square dots).

The effect of the ortho-normalization in a correlated orbital varies depending on the functional. SDFT+U implemented using the un-normalized projector as adopted in VASP (circular open dots and dashed lines, also shown in Ref.~\onlinecite{Park:14}) moderately reduces the critical pressures (favoring the $\delta a>$0 region)
compared to the same SDFT+U implemented using the ortho-normalized projector (circular filled dots and solid lines).  In contrast, DFT+U with the ortho-normalized projector (square filled dots and solid lines) more substantially moves the critical line toward the structural phase with $\delta a$=0 ($Pbnm$) for the heavy rare earths.

DFT+U implemented using the MLWF basis set (blue diamond dot and solid line, also shown in Ref.~\onlinecite{Park2014235103} with $u$=5eV and $j$=1eV of Slater-Kanamori parametrization)  also produces a similar phase boundary to the orthonormalized projector DFT+U calculation compared to other methods. Therefore, we deduce that Wannier and orthonormalized projectors yield similar behavior.

\begin{figure}[!htbp]
\includegraphics[scale=0.45]{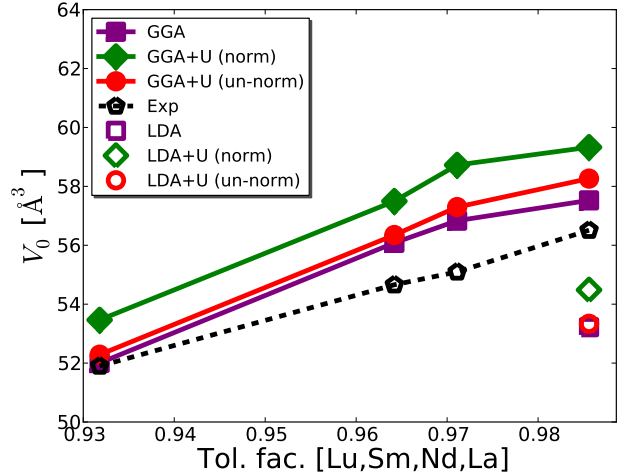}
\caption{(Color online)
Equilibrium volume $V_0$ calculated using SGGA (filled symbols) and LSDA (open symbols)  for $R$NiO$_3$ with $R$=La, Sm, Nd and Lu using DFT+U (square dots) and SDFT+U. For SDFT+U, ortho-normalized projector (green diamond) and  un-normalized projector (red circle) calculations are compared.  The experimental volumes at the ambient pressure are depicted  by a black dashed line with open pentagonal dots. 
LDA(+U) results are shown only for LaNiO$_3$ since the pseudo potentials for other rare earth ions except La are not available.
\label{fig:Eq_vol}}
\end{figure}

The  compression (vertical axis) of the structural phase diagram in Fig.$\:$\ref{fig:phased_sp} is defined as the relative change of volume compared to the equilibrium volume $V_0$ computed within each theoretical method. $V_0$  is determined as the volume of the minimum energy from an energy vs volume curve  and the atomic positions at each volume are obtained by minimizing the inter-atomic forces. Results are displayed in Fig.$\:$\ref{fig:Eq_vol}
First, we discuss results obtained using  pure SDFT within the LSDA and SGGA approximations. The calculated $V_0$ with GGA exchange-correlation functional (filled symbols) is larger than the experimental one (dashed line),  while the $V_0$ obtained with the LDA exchange-correlation functional (open symbols) is smaller than the experimental value. This behavior is well known in the DFT literature. 

We next observe that the calculated $V_0$ values computed using SDFT+U are rather sensitive to the ortho-normalization of correlated orbitals. The orthonormalized projectors lead to substantially larger equilibrium volumes than the un-normalized projectors, leading to  better agreement with experiment for LDA based functionals and worse agreement for GGA, though in neither method is the agreement particularly good.

\section{Conclusion \label{sec:conclusion}}

We studied different formulations of DFT+U and SDFT+U in the context of total energy calculations in the rare-earth nickelates.  The correlated subspace was constructed in three different ways: maximally localized Wannier functions, orthonormalized projectors, and non-orthonormalized projectors.  We computed the Ni-O bond length difference $\delta a$ as a function of pressure, the structural phase diagram describing the transition between the bond-disproportionated structure ($\delta a>$0) and the no  bond-disproportionated structure ($\delta a$=0) as functions of pressure and the rare-earth ions, and also the equilibrium volume.

SDFT+U and DFT+U show qualitatively different behavior in some circumstances.  In particular, the SGGA+U results of the structural transition in the rare-earth nickelates show a re-entrant transition with pressure, and this is not observed in GGA+U calculations that are performed with a reasonable on-site exchange $J=1eV$. However, increasing $J$ to 2eV in GGA+U produced similar results to SGGA+U, implying that the SGGA spin-polarized exchange correlation functional results in a large effective on-site exchange.  SDFT+U based on the LSDA exchange-correlation functional results in a more reasonable effective $J$, meaning that LSDA+U results using $J$=0 are similar to LDA+U with $J\sim$1eV.  The reentrant transition at negative pressure does not occur within DFT+U calculations using the LDA functional for $J\le$2eV.  Our results suggest that there is no need to use an on-site exchange when performing SDFT+U (ie. set $J=0$), and this is effectively equivalent to using the approach of Dudarev et al\cite{Dudarev19981505}.  Additionally, our results imply that the  SGGA should be used with caution given its overemphasis of local exchange. More generally, DFT+U with an appropriately chosen $J$ can largely recover the qualitative behavior of SDFT+U.

We demonstrated that orthonormalized projectors behaved rather similarly to MLWF near ambient pressure; although notable differences are evident for NdNiO$_3$ and LaNiO$_3$ under large pressures. Additionally, the un-normalized projector, as implemented in VASP, can lead to  notably different results, especially at high pressures.  Within SDFT+U, the equilibrium volumes are substantially increased when computed using the ortho-normalized orbitals compared to un-normalized orbitals.

Given that (S)DFT+U is equivalent to (S)DFT+DMFT when the DMFT quantum impurity problem is solved within Hartree-Fock, we expect the general findings of this study to be applicable to (S)DFT+DMFT as well.
Given our finding that DFT+U with an appropriately chosen $J$ can largely recover the qualitative behavior of SDFT+U, our work supports the long held tradition of basing dynamical mean field extensions on DFT theories rather than SDFT theories. This is particularly important in cases where the dynamical effects of $J$ are crucial.
Our results have broad implications for the application of SDFT+U, given that in the nickelates the choice of SDFT functional leads to dramatic differences in the effective on-site exhcange interaction. A very similar conclusion was reached in a study of the spin crossover molecule Fe(phen)$_2$(NCS)$_2$\cite{JiaChen:2015}.

{\it Acknowledgements} AJM acknowledges support from the Basic Energy Sciences Division of the US Department of Energy under grant ER-046169. HP and CAM acknowledge support from FAME, one of six centers of STARnet, a Semiconductor Research Corporation program sponsored by MARCO and DARPA. This research used resources of the National Energy Research Scientific Computing Center, a DOE Office of Science User Facility supported by the Office of Science of the U.S. Department of Energy under Contract No. DE-AC02-05CH11231.

\vspace{0.05in}

\appendix

\section{Computational details \label{Appendix:sec:Theory-4}}

The (S)DFT+U formalism is implemented in VASP using the projector functions $|\tilde{p}\rangle$ as a correlated orbital set.
The Hamiltonian for (S)DFT+U is given by Eq.$\:$\ref{eq:H_PAWU}, and the total energy and force equations are also derived in Eq.$\:$\ref{eq:E_PAWU} and Eq.$\:$\ref{eq:F_PAWU}.
However, the VASP implementation adopt the un-normalized density matrix $n^{\tau\sigma}$ as Eq.$\:$\ref{eq:PAW}.
In the current paper, we compare the VASP implementation to  the ortho-normalization of $n^{\tau\sigma}$ to give rise to $\bar{n}^{\tau\sigma}$ as derived in Eq$\:$\ref{eq:norm}.
Fortunately, the VASP implementation provides both DFT+U (LDAUTYPE=4) and SDFT+U (LDAUTYPE=1) methods.

For performing the summation of $\mathbf{k}$ points in the Brillouin zone,
we used the tetrahedron method~\cite{Blochl:94}.
When using projector correlated orbitals, a $\mathbf{k}$-point mesh of $6\times6\times6$ (for the $Pbnm$ and $P2_1/n$ structures) and $8\times8\times8$ (for the $La$NiO$_3$ $R\bar{3}c$ and $R3$ structures) are used with an energy cutoff of 600eV. 
When using the Wannier functions as correlated orbitals, $\mathbf{k}$ points meshes of size $10\times10\times10$ is used for $Pbnm$ and $P2_1/n$, while $16\times16\times16$ for $R\bar{3}c$ and $R3$.

\bibliography{main}

\end{document}